\documentclass[journal]{IEEEtran}

\ifCLASSINFOpdf
\else
\fi
\usepackage{amssymb}
\usepackage{xcolor}
\usepackage{graphicx}
\usepackage{bm}
\usepackage{amsmath}
\usepackage{mathtools}
\usepackage{algorithm}
\usepackage{caption} 
\usepackage[noend]{algpseudocode}
\usepackage[
backend=biber,
style=numeric-comp,
sorting=ynt,
hyperref=true,
maxnames=10
]{biblatex}
 
\usepackage[utf8]{inputenc}
\usepackage[english]{babel}
 
\usepackage{amsthm}
 
\theoremstyle{remark}
\newtheorem*{remark}{Remark}

\bibliography{egbib.bib}

\DeclareMathOperator*{\argmax}{\arg\!\max}
\DeclareMathOperator*{\argmin}{\arg\!\min}

\begin{document}
%
\title{A scalable algorithm for identifying multiple sensor faults using disentangled RNNs}
%
%
%

\author{David~Haldimann,
        Marco~Guerriero,~\IEEEmembership{Senior Member},
        Yannick~Maret,~\IEEEmembership{Senior Member},
        Nunzio~Bonavita,
        Gregorio~Ciarlo,
        Marta~Sabbadin
\thanks{David~Haldimann, Marco Guerriero and Yannick Maret
are with ABB Switzerland Ltd, Corporate Research Center, Baden, Switzerland.}
\thanks{Nunzio~Bonavita, Gregorio~Ciarlo and Marta~Sabbadin are with ABB SpA, Industrial Automation Measurement Analytics, Genova, Italy.}
}

%
%

\markboth{IEEE Transactions on Neural Networks and Learning Systems}%
{Shell \MakeLowercase{\textit{et al.}}: Bare Demo of IEEEtran.cls for IEEE Journals}
%



\maketitle

\begin{abstract}

The problem of detecting and identifying sensor faults is critical for efficient, safe, regulatory-compliant and sustainable operations of modern systems.
Their increasing complexity brings new challenges for the Sensor Fault Detection and Isolation (SFD-SFI) tasks.
One of the key enablers for any SFD-SFI methods employed in modern complex sensor systems, is the so-called analytical redundancy, which is nothing but building an analytical model of the sensors observations (either derived from first principles or identified from historical data in a data-driven fashion).
In a nutshell, SFD amounts to generate and to monitor residuals by comparing the sensor measurements with the model predictions with the idea that the faulty sensors will result in large residuals (i.e. the defective sensors generate measurement that are inconsistent with their expected behavior represented by the model).
In this paper we introduce a disentangled Recurrent Neural Network (RNN) with the objective to cope with the \textit{smearing-out} effect, i.e. the propagation of a sensor fault to the non-faulty sensors resulting in large misleading residuals.
Moreover, the introduction of a probabilistic model for the residual generation allows us to develop a novel procedure for the identification of the faulty sensors. 
The computational complexity of the proposed algorithm is linear in the number of sensors as opposed to the combinatorial nature of the SFI problem.
Finally, we empirically verify the performances of the proposed SFD-SFI architecture using a real data set collected at a petrochemical plant.

\end{abstract}


\IEEEpeerreviewmaketitle

\section{Introduction}
\subsection{Background, Motivation, State of the Art}
Modern industrial process plants are large scale, highly complex systems continuously measured and monitored by a large number of sensors to ensure product quality and efficient and safe operations \cite{fortuna2007soft}.
Accurate fault detection and diagnosis are of the utmost importance to minimize downtime, increase the safety of the plant operations and meet the increasingly stringent safety and environmental regulation requirements \cite{jahnke2000continuous, ciarlo2017assessment,angelosante2018sensor}.
Any component of an industrial process plant can be susceptible to a fault \cite{isermann1997trends}. Among different types of faults, such as sensor faults, actuator faults and process faults, the focus of this paper is on the faults affecting the sensors. 
The complexity of modern industrial plants equipped with a very large number of sensors, poses new challenges on the human capability of manually validating the sensor operational status using traditional Key Performance Indicators (KPIs) and dashboards.
A new level of automation is required, in order to generate and handle the enormous number of different hypotheses related to potential root-causes, in a much more accurate, efficient and scalable way than it is done today \cite{zhang2017kpi}.
Therefore, in the new industry 4.0 context  \cite{gilchrist2016industry}, the need of designing monitoring systems that can scale to thousands, even millions of sensors, is becoming increasingly urgent.

A classical and recurring example of sensor fault is given by a sensor producing biased measurements \cite{webster2016measurement}.
In order to safeguard industrial plants according to operational requirements, sensor faults have to be detected, isolated and whenever possible, the sensor values have to be reconstructed \cite{Quevedo2017}.
The first two procedures, which are named Sensor Fault Detection (SFD) and Sensor Fault Isolation or Identification (SFI), are in the scope of this work.
SFD is all about determining whether a fault in one sensor or multiple sensors has occurred, while SFI concerns with identifying (or isolating) the subset of faulty sensors. 

The approaches to SFD-SFI typically fall into three categories; namely, data-driven, analytical-based, and knowledge-based. The data-driven model is derived directly from historical sensor data using either statistical or machine learning techniques \cite{qin2012survey,naderi2018data, yin2014data}. Unlike the data-driven approach, the analytical approach uses mathematical models often derived from first principles \cite{isermann2005model,simani2000}.
It is often difficult, if not impossible, to apply the analytical approach to large-scale systems, because detailed models are required in order for the approach to be effective. Such models are very expensive to obtain for large-scale systems given all the interactions associated with the process and the multiple heterogeneous sensors \cite{isermann1994integration}.
Finally, the knowledge-based approach relies upon qualitative models, based on causal analysis, expert systems, to develop sensors monitoring measures \cite{da2012knowledge}. 

All three approaches have pros and cons, so that no single approach is the best for all applications.
In this paper we follow the \textit{data-driven} approach.
Within this category, Principal Component Analysis (PCA) \cite{harkat2006improved,li2001consistent}, Fisher Discriminant Analysis (FDA) \cite{du2008multiple}  analysis and Partial Least Squares (PLS) \cite{chiang2000fault} are the most widely used classical statistical techniques.
Machine learning algorithms, with neural networks playing the role of the master tool, are also popular approaches \cite{ntalampiras2014fault, reppa2013adaptive, jabbari2007application, maki1997neural, vemuri1998neural, marcu1997robust}.
The key ingredient for the data-driven approach is the availability of representative historical data from which one can learn a model for the sensors observations.
The learning problem is classified into two categories: \textit{unsupervised} and \textit{supervised} learning \cite{friedman2001elements}, depending on whether the training data set includes or not the normal sensor states and faults, both labeled as such.
Our working hypothesis is that only sensors measurements in the normal state are available, leading us to search a potential solution to the SFD and SFI problems, in the space of unsupervised approaches. 

In the analytical-based approach, SFD can be performed by comparing the behavior of the mathematical model describing the sensors measurements, with that of the actual sensors observations under analysis. Deviations between the actual sensors measurements and the sensor values calculated by the model, might indicate that a fault has occurred. This form of redundancy is called analytical redundancy \cite{chow1984analytical}. Analytical methods that use \textit{residuals} as decision statistics for the SFD and SFI problems, are commonly referred to analytical redundancy methods \cite{gertler1991analytical}.
Ideally, the residuals will be large when the sensor faults are present, and small in the fault-free case\footnote{The small residuals when there are no faults, are due to the presence of noise, and/or modeling errors.}.
When dealing with multiple sensors, one is faced with a vector of residuals, one for each sensor. Therefore the norm of this vector is used to determine the occurrence of a fault. 

Instead, SFI is performed by examining the \textit{contribution} of each individual sensor residual to the global one \cite{alcala2009reconstruction}.
Unfortunately the contribution plot suffers from the \textit{smearing-out} effect \cite{westerhuis2000generalized}, which is the influence of faulty sensors on the contribution of non-faulty sensors, leading to the miss-classification of the faulty sensors.
The smearing-out effect is due to the predicted value of one sensor being the result of a combination (that can be highly non-linear) of all the other sensors inputs to the prediction model. The interaction (entanglement) of faulty sensors and non-faulty ones embedded in the analytical model is the fundamental cause of the smearing-out phenomenon.

\subsection{Main Contributions}

We borrowed the idea of using the residuals as decision statistics from the analytical-based literature, with the important difference that we learn the model from the data instead of using first principles analysis.
More specifically, we decide to use a Recurrent Neural Network (RNN) \cite{goodfellow2016deep} to capture the spatial-temporal relationships among the different sensors measurements during the fault-free case and use a residual analysis to solve both the SFD and SFI problems.
In the recent years, RNNs have been successfully applied to anomaly detection for multiple time series \cite{Malhotra2015LongST, Obst2014, Chauhan2015, Malhotra2016}, where a prediction model based on RNNs is used to reconstruct the ``normal" multiple time-series behavior, and thereafter the reconstruction error is used to detect anomalies.

The main contributions of this paper can be summarized as follows:
\begin{itemize}
    \item We introduce a new RNN architecture, that we call \textit{disentangled} RNN, with the objective of neutralizing the smearing-out effect that plagues SFI.
    \item We build a probabilistic characterization of the residuals to guide the design choices for the SFD and SFI systems and to characterize their performances too.
    \item Using the aforementioned probabilistic framework, we introduce a novel greedy algorithm, called \textbf{GreedyIso}, for identifying multiple sensor faults, with computational complexity that is \textit{linear} in the number of sensors to be monitored.
\end{itemize}
We would like to emphasize that our idea of using disentanglement against the smearing-out effect, can be applied with other models as well. Disentanglement acts as a regularization term that can be used by any machine learning model and it is not therefore restricted to the RNN class of models. The same holds also for the other results of the papers.

\subsection{Notation}
Throughout the paper, vectors and lists are expressed
in bold face letters, while matrices are expressed in capital
bold face letter. Here we denote with $S$ the number of sensors. 
We denote with $\bm{x}_{t}:=[x^1_t,\ldots,x^S_t]^\top\in
\mathbb{R}^S$ (with $^\top$ we denote the transposition symbol) the stacked vector of the sensors measurements at time $t$. We will refer to it with the generic term "signal" throughout the rest of the paper.
The complementary hypotheses of \textit{no} sensor faults and \textit{at least one} sensor fault are denoted with $\mathcal{H}_0$ and $\mathcal{H}_1$, respectively.
The symbols $\|\cdot\|_2$ and $\|\cdot\|_1$ denote the $L_2$ norm of a vector and the element-wise L1 matrix norm, respectively.

\subsection{Assumptions}
Calibration errors are probably the key source of many faults manifesting themselves as a bias or a drift in the sensor readings. Moreover, faults can be classified based upon the temporal persistence as \textit{permanent faults} that are continuous and stable in time; \textit{intermittent faults} and \textit{transient faults}.
In this study we focus our attention on the permanent faults characterized by continuity of the fault signature occurrence leading to an observable pattern that we can learn over time \cite{ni2009sensor}.
A quiet general model for the multiple faults case \cite{gorinevsky2015fault}, \cite{li2014sensor} is given by:
\begin{equation}
\bm{\Tilde{x}}_{t}(\bm{\Delta}) = \bm{x}_{t} + \bm{\Delta}
\label{eq:sensor_fault}
\end{equation}
where a bias vector $\bm{\Delta} = [\Delta^1,\ldots,\Delta^S]^\top\in \mathbb{R}^S$ is added to the nominal sensors vector $\bm{x}_{t}$ to generate the faulty vector $\bm{\Tilde{x}}_{t}$.

\subsection{Paper Organization}

The remainder of this paper is organized as follows. In
Section \ref{sec:rnn}, we provide an overview of RNNs on how they can be used to build the model for the sensors observations. 
In Section \ref{sec:SFD}, we introduce the SFD problem with the probabilistic framework of the residuals.
Section \ref{sec:SFI} is dedicated to the description of the SFI problem, while Section \ref{sec:disent} introduces the disentangled RNN.
Section \ref{sec:greedyApproach} introduces the GreedyIso algorithm and its variant, called GreedyIsoSparse.
In Section \ref{sec:two_stage} we examine two different architectures for coupling SFD with SFI. 
Section \ref{sec:num_results} assesses the performances of the proposed algorithm in a combination of simulated and real
scenarios. Section \ref{sec:conclusions} concludes the paper.

\section{Recurrent Neural Networks}
\label{sec:rnn}

\subsection{Some history}
A Recurrent Neural Network (RNN) is a special type of neural network that can capture temporal dependencies. Modeling temporal data is critical in most real-world applications, since signals like speech and video have time-varying properties and are characterized by having dependencies across time \cite{graves2014towards}, \cite{liang2015recurrent}.
The first attempt to add memory to neural networks were the Time Delay Neural Networks (TDNNs) \cite{waibel1995phoneme}, in which inputs from past time steps are introduced to a regular Feed Forward Neural Network (FFNN). This has the advantage of clearly allowing the network to look beyond the current time step, but also introduces a disadvantage since temporal dependencies are limited to the size of the chosen time window.
In contrast to this approach, simple RNNs, also known as Elman networks \cite{elman1990finding} and Jordan networks \cite{jordan1997serial}, were using an internal state called ``memory". However in the early 90's, it was recognized that these networks suffer from the vanishing gradient problem \cite{schmidhuber1992learning} in which contributions of information decayed geometrically over time. Capturing relationships that spanned more than eight or ten steps back was practically impossible. In the mid 90's, Long Short-Term Memory cells, or LSTMs in short, were invented to address this very problem \cite{hochreiter1997long}. The key novelty in LSTMs was the idea that arbitrary temporal dependencies can be captured by using gates that control the flow of the internal state variables in and out of the network. Variations on LSTMs such as Gated Recurrent Units (GRUs) \cite{cho2014properties}, further refine this theme, and nowadays, represent another mainstream approach to implement RNNs.

\subsection{RNNs used for predicting multiple sensor observations}

The objective of the RNN is to capture both the correlation among sensors and the inherent temporal dependency. This is achieved by learning a non-linear function $\bm{g^{\bm{\theta}}}$ (parametrized by the training vector $\bm{\theta}$) that maps the sequence of past sensors observations to a single output sensors observations at time $t$.
More specifically, we have that the predicted sensors values at time $t$ are given by:
\begin{equation}
\bm{\hat{x}}_{t} = \bm{g^{\bm{\theta}}}(\bm{\hat{x}}_{t-1}, \bm{x}_{t-1}).
\label{lab:pred}
\end{equation}

Unlike the FFNNs, the RNNs do not require a pre-specified window of previous time steps to use as input variables to predict the sensors values at the next time period. They can remember what they have seen in the past even if trained with a single training sample at a time, through the cell state capturing the memory of the system as described by the recurrence relation in (\ref{lab:pred}). The ``right" size of the window of inputs to remember is learned from the data.

\begin{figure*}
\begin{center}
\includegraphics[width=1.0\linewidth]{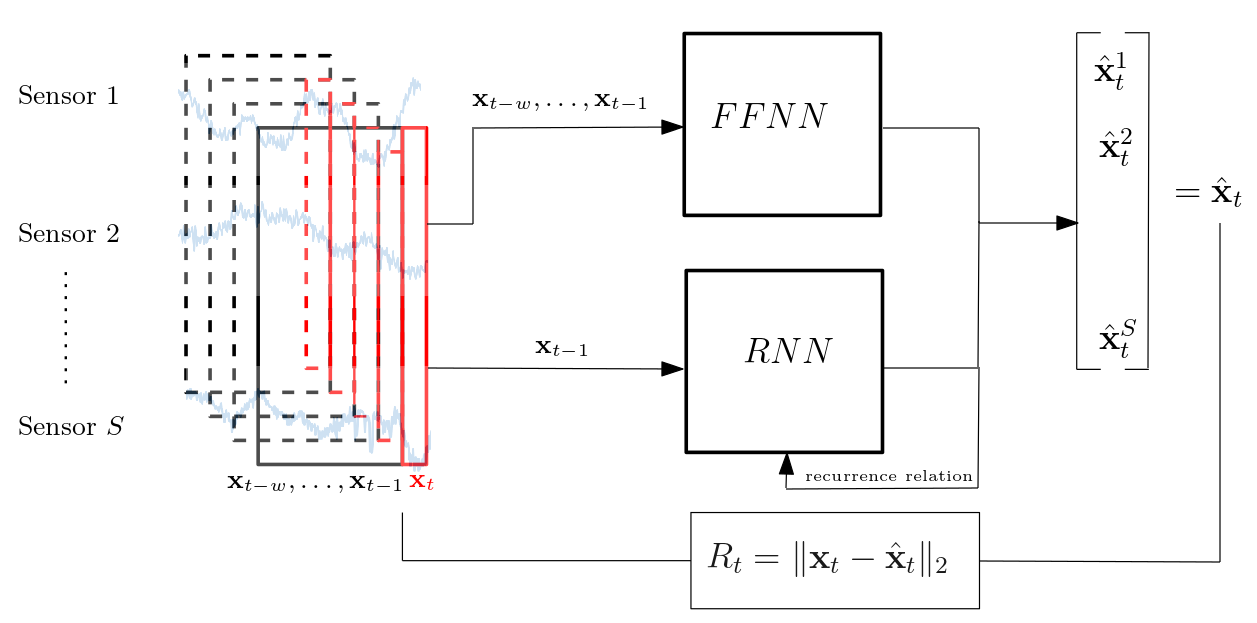}
\end{center}
   \caption{Pictorial illustration of both the RNN and FFNN architectures deployed to generate the residual $R_t$ defined in Section \ref{sec:SFD}.
   Using the FFNN approach, the predicted signal $\hat{\bm{x}}_{t}$ at time $t$ is computed from a fixed window of size $w$ of the previous terms, which is $[\bm{x}_{t-1},...,\bm{x}_{t-w}]$, while the RNN uses the previous term $\bm{x}_{t-1}$ and its internal memory (via the recurrence relation) to predict the sensors values at time $t$. Then the residual $R_t$ is defined as the $L_2$ norm of the vector difference between the actual sensors measurements $\textcolor{red}{\bm{x}_{t}}$ and their expected behavior according to the model, which is given by the signal $\hat{\bm{x}}_{t}$.}
\label{fig:shortGraph}
\end{figure*}

On the other hand, a sliding window is necessary for multiple time series forecasting using FFNNs, because FFNNs require a fixed size input and do not have memory. The FFNN aims at learning the following non-linear function $\bm{h^{\bm{\xi}}}$ (parametrized by the training vector $\xi$):

\begin{equation}
\bm{\hat{x}}_{t} = \bm{h^{\bm{\xi}}}([\bm{x}_{t-1},...,\bm{x}_{t-w}])
\label{lab:pred_ffnn}
\end{equation}
with $w$ being the length of the look back period.
A visualization of the proposed architecture can be seen in Figure~\ref{fig:shortGraph}. 
In the numerical section, a comparison between the performance of the RNN and FFNN is provided.

Given the training set $\bm{X}:=[\bm{x}_1,\ldots,\bm{x}_N] \in \mathbb{R}^{S\times N}$ where $\bm{x}_t$ represents the sensors measurements at time $t$ and $N$ is the length of the training set, one can define the following loss function:

\begin{equation}
\bm{\mathcal{L}}_{MSE} = \frac{1}{N} \sum\limits^{N}_{t=1} \| \bm{x}_{t} - \bm{\hat{x}}_{t}^2\|_2
\label{eq:loss}
\end{equation}
where $\bm{\hat{x}}_t$ is the predicted signal using either RNN or FFNN as per equation (\ref{lab:pred}) or (\ref{lab:pred_ffnn}).
The problem of learning amounts to estimating the parameter vector $\bm{\theta}$ that minimizes $\bm{\mathcal{L}}_{MSE}$.
The parameter optimization can be implemented with stochastic gradient decent (SGD) or second order gradient algorithms \cite{goodfellow2016deep}. In this work we use the Adam algorithm \cite{kingma2014adam} as modified stochastic gradient descent approach.


\section{SFD: Problem formulation}
\label{sec:SFD}

In designing the SFD system we do \textit{not} assume any knowledge about the generative model of the sensors observations. This prevents us from building a probabilistic model for the sensors. However, we do assume the existence of a \textit{training set} $\bm{X}$.
The data $\bm{X}$ are observed during the data collection period and they capture a wide variety of operating points of the process. Clearly, if an operating point is not observed during the measurement campaign, certain algorithms might detect the presence of faulty sensors even if the process is simply working at an operating point different from those examined during the data collection period.
We also assume that the data set $\bm{X}$ is not polluted by persistent faulty sensor.
The goal of SFD is to detect the presence of faulty sensors by checking whether a new sensors measurements vector $\bm{x}_{t}$ at time $t$ is a novelty \cite{Markou2003}.

Given the residual vector $\bm{r}_t$ defined as
\begin{equation}
\bm{r}_t = \bm{x}_t - \bm{g^{\bm{\theta}}}(\bm{\hat{x}}_{t-1}, \bm{x}_{t-1}).
\label{lab:res}
\end{equation}
The norm of the residual vector $\bm{r}_t=[r^1_t,...,r^S_t]$ for $S$ sensors is defined as:
\begin{equation}
R_t = \|\bm{r}_t\|_2
\label{lab:res_norm}
\end{equation}
In the faulty-free case (we denote the hypothesis of no-faults with the symbol $\mathcal{H}_0$), one would expect $R_t$ being relatively small as it only captures the training error but not the sensor fault signature.
On the other hand, when a fault in one or more sensors is present (we denote the hypothesis of \textit{at least one} sensor fault with the symbol $\mathcal{H}_1$), $R_t$ is expected to be larger due to the effect of the fault signature inflating the RNN model mismatch.
At each time $t$, the binary hypothesis testing problem is:

\begin{equation} \label{eq:binary_hypo}
\begin{split}
\mathcal{H}_0:  \textit{ } & R_t \approx \textit{small}
\\
\mathcal{H}_1:  \textit{ } & R_t \approx \textit{large}
\end{split}
\end{equation}
The mathematical rigour of the problem in  (\ref{eq:binary_hypo}) is intentionally loose to reflect the lack of statistical model for the residuals. 
The natural approach to discriminate between the two hypotheses is to monitor $R_t$ by comparing it against a threshold $\gamma$. A decision in favor of $\mathcal{H}_1$ will be made whenever $R_t$ exceeds $\gamma$. 
Mathematically, this amounts to define the decision binary  variable denoted with the indicator function $D_t=I(R_t>\gamma)$\footnote{The indicator function $I(R_t>\gamma)$ is equal to 1 if the event $R_t>\gamma$ is true, 0 vice versa.}.
The decision $D_t \in \{0,1\}$ at time $t$, takes the value 1 when there is a detection of a faulty situation; otherwise, it takes the value 0.

Because of the unsupervised nature of the problem (only the training data with no faults are available), we can estimate the probability density function (pdf) of $R_t$ only under $\mathcal{H}_0$ \footnote{The pdf of $R_t$ under $\mathcal{H}_0$ can be estimated via a kernel density estimator relatively easily due to its uni-variate nature \cite{wand1994kernel}.}. This gives us the way to compute (via numerical integration) the probability of false alarm $p_{fa}$\footnote{Note that $p_{fa}$ is independent of time since it is assumed that the process $R_t$ is stationary \cite{papoulis2002probability}.}as:

\begin{equation}
p_{fa} = P\{R_{t} > \gamma|\mathcal{H}_0\} = \int\limits_{ \gamma}^{\infty}f(R_{t}|\mathcal{H}_0)dR_{t}
\label{lab:pfa}
\end{equation}
where with $P\{\cdot\}$ and $f(R_{t}|\mathcal{H}_0)$ we denote the probability of an event and the pdf of $R_t$ under $\mathcal{H}_0$, respectively.
In the same way, we can define the probability of detection $p_d$ as:
\begin{equation}
p_{d} = P\{R_{t} > \gamma|\mathcal{H}_1\} = \int\limits_{ \gamma}^{\infty}f(R_{t}|\mathcal{H}_1)dR_{t}
\label{lab:pd}
\end{equation}
The pdf $f(R_{t}|\mathcal{H}_1)$ is obviously unknown (and it cannot be learned from data since sensors faulty data are not available). Nevertheless, equation (\ref{lab:pd}) would become very useful for:
\begin{itemize}
    \item characterizing the performance of the SFD algorithm and its variants (with and without disentanglement).
    \item guiding in a principled way the greedy algorithm described in Section \ref{sec:greedyApproach}.
\end{itemize}

\begin{figure*}
\begin{center}
\includegraphics[width=0.8\linewidth]{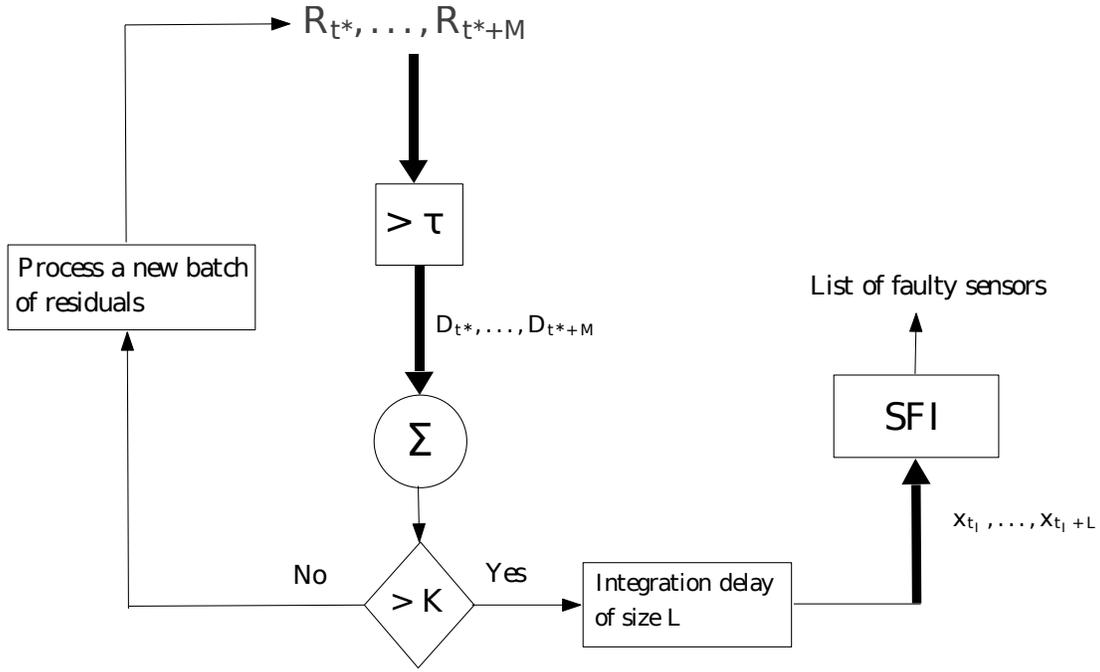}
\end{center}
   \caption{The SFD system processes a batch of size $M$ composed of individual binary variables obtained after thresholding each individual residual in the batch (the thick lines in the diagram denote the multiple input/output information flowing in and out of the different processing blocks). The SFI is then activated (after a delay of size $L$) to output the list of faulty sensors by processing the new sensors measurements in the batch of size $L$.}
\label{fig:sfd_sfi}
\end{figure*}

\subsection{Decision Fusion}

The focus of this paper is on detecting \textit{persistent} sensor faults. This suggests us to integrate the decision statistics over a certain time interval to increase the overall detection performance of the SFD system and also its robustness against intermittent and transient faults at the price of introducing a delay in making the final decision.
The integration procedure is implemented in batch mode as opposed to streaming mode where the data are processed sequentially one point at a time or on rolling windows \cite{siegmund2013sequential, glaz2012scan, basseville1993detection, xie2013sequential, Guerriero2015changedetection, guerriero2009distributed}.
Within the class of batch processing, two integration strategies
can be envisaged, with the second being adopted in this work: i) \textit{Data fusion} \cite{varshney2012distributed} wherein the raw residuals $R_t$ are combined to make the final assessment on the presence of the sensors faults; ii) \textit{Decision fusion} wherein multiple individual decisions $D_t$ are fused together to produce the final decision \cite{varshney2012distributed}.

We are now in the position to detail the operations of the SFD system.  Once activated at time $t^*$, the SFD system will start to store the individual decision $D_t$ into a decision vector $\bm{D}=\{D_t: t=t^*,...,t^* + M\}$. The final decision rule $\mathcal{D}(K,M)$ is then given by:

\begin{equation}
\mathcal{D}(K,M) = 
\begin{cases}
\text{decide in favor of } \mathcal{H}_1 &\text{if } \sum\limits^{t^*+M}_{t=t^*}D_t \geq K, \\
\text{decide in favor of } \mathcal{H}_0  &\text{if } \sum\limits^{t^*+M}_{t=t^*}D_t < K.
\end{cases}
\end{equation}
The decision rule is also known as $K$-out-of-$M$ rule \cite{viswanathan1997distributed}, \cite{bar2011tracking} for which a fault in one or more sensors is declared when there are least $K$ individual detections in the decision vector of size $M$.

The system level probability of false alarm associated to the detector $\mathcal{D}(K,M)$ is given by:

\begin{equation}
P_{fa} = P\{\sum\limits^{t^*+M}_{t=t^*}D_t \geq K|\mathcal{H}_0\}
\label{lab:pfa_global}
\end{equation}
It is not difficult to show that, under the hypothesis $\mathcal{H}_0$, the total number of detection $ \sum\limits^{t^*+M}_{t=t^*}D_t$ approximately \footnote{The approximation follows from the fact that the variables $D_t$ are identically distributed Bernoulli random variables with parameter $p_{FA}$ but not independent since the residuals $R_t$ are dependent over time due to the intrinsic temporal nature of the RNNs. For the interested reader, \cite{glaz2012scan} includes several approximations for sums of dependent random variables.} follows a Binomial distribution $\mathcal{B}(M, p_{fa})$ with parameters $M$ and $p_{fa}$ \cite{papoulis2002probability}.
Therefore, for a given threshold $K$, the probability of false alarm $P_{fa}$ can be approximated as follows:
\begin{equation}
P_{fa} \approx \sum\limits^{M}_{i=\lceil K \rceil}\binom{M}{i}p^i_{fa}(1-p_{fa})^{M-i}
\label{lab:Pfa}
\end{equation}
where the symbol $\lceil z \rceil$ means the ceiling of $z$, i.e. the smallest integer greater than or equal to $z$.

Note that for hypothesis testing problems involving discrete distributions, it is usually not possible to design a parameter $K_{\alpha}$ to achieve a probability of false alarm $P_{fa}$ exactly equal to a significance level $\alpha$, where $\alpha$ is a some prescribed value that is specified by the user according to the SFD system requirements. 
The above issue is addressed by resorting to a \textit{randomized test} \cite{poor2013introduction}.
We define $\alpha_1$ and $\alpha_2$ as follows:
\begin{equation} \label{eq:randomized_test}
\begin{split}
P\{\sum\limits^{t^*+M}_{t=t^*}D_t \geq K_{\alpha}|\mathcal{H}_0\} & \approx \alpha_1 < \alpha
\\
P\{\sum\limits^{t^*+M}_{t=t^*}D_t \geq K_{\alpha} -1 |\mathcal{H}_0\} & \approx \alpha_2 > \alpha
\end{split}
\end{equation}
then we use a "coin-flip" decision with probability
\begin{equation} \label{eq:coin_flip}
p_{flip} = \frac{\alpha - \alpha_1}{\alpha_2 - \alpha_1}
\end{equation}
whenever $\sum\limits^{t^*+M}_{t=t^*}D_t = K_{\alpha}$.
To obtain the complete set of the SFD system level performance measures, we also need to compute the corresponding $P_D$ value, which is given by:
\begin{equation}
P_{D} \approx  \sum\limits^{M}_{i=K_{\alpha}}\binom{M}{i}p_{d}^i(1-p_{d})^{M-i}
\label{lab:Pdetect}
\end{equation}

Throughout the rest of the paper we will make use of $P_D$ as benchmark performance metric for evaluating both the SFD and SFI algorithms. 

\begin{remark}
The user of the SFD system will have to design the different parameters $P_{fa}$, $p_{fa}$, $K$ and $M$ while considering the interplay among them.
For instance one would want to set $\alpha=0.1$ resulting in 10 false alarms over $100$ different batches of data, on average, where each batch is composed of $M$ sensors measurements. This requirement automatically imposes a condition on $p_{fa} \leq \frac{\alpha}{M}$ that helps set the threshold $\gamma$.
\end{remark}

\begin{figure*}
\begin{center}
\centerline{\includegraphics[width=1.3\linewidth]{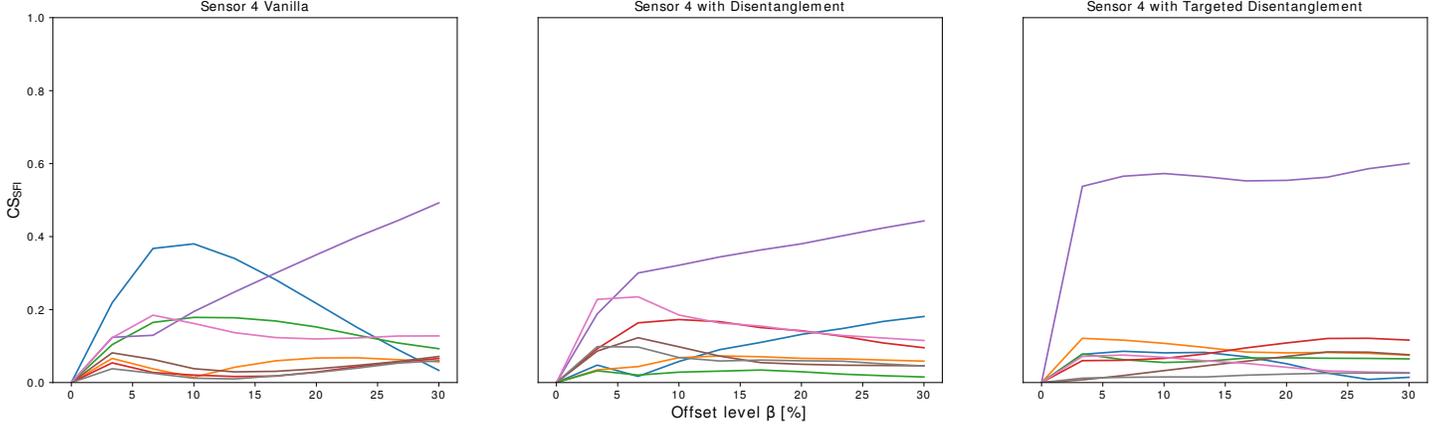}}

\end{center}
   \caption{Comparing contribution plots in order to study the impact of disentanglement in challenging situations where the sensor fault signature is small and hard to detect. The contribution scores $\{CS_{SFI}(s)\}_{s=1}^{8}$ for $S=8$ sensors are plotted against the increasing offset inserted only into one sensor, which is the one depicted with the purple color.
   More specifically, for this toy example we have that $\bm{\Delta} = [0,\ldots,\Delta^i,\ldots,0]^\top \in \mathbb{R}^8$ is a $8$-dimensional  vector with all zero entries except in the position $i$ corresponding to the purple sensor where the entry is equal to $\Delta^*$. The $x$-axis shows the offset level $\beta$ in percentage meaning that $\Delta^*=\beta \bar{x}^{i}$, with $\bar{x}^{i} = \frac{1}{N}\sum_{n=1}^{N}x^{i}_{n}$ being the expected value of the no-faulty observations for the $i$-th sensor in the training set.
   The left plot shows the contributions generated using no disentanglement. For the middle plot the disentanglement regularization introduced in Section \ref{sec:disent} is used. Lastly, the plot on the right shows the contribution when using targeted disentanglement as described in Section \ref{sec:tarDisent}.}
\label{fig:contribPlot}
\end{figure*}

\section{SFI: Problem Formulation}
\label{sec:SFI}
After a fault has been declared at time $t_{fault} = t^* + M$, the next step is to \textit{isolate} the faulty sensors.
The input to the SFI system is a set of consecutive sensors measurements after the time $t_{fault}$, which is $\{\bm{x}_t\}_{t=t_{I}}^{t_{I} + L}$, where $t_I$ and $L$ are the activation time and the size of integration time interval of the SFI processing block. The logical condition $t_I > t_{fault}$ ensures that only the sensors measurements that are deemed faulty, are fed to the SFI unit, which is responsible to output the set of faulty sensors out of the $S$ sensors. Moreover, the SFI integration time $L$, although it introduces a delay, helps make the process of estimating the faulty sensors, more accurate.
A graphical overview of the SFD-SFI pipeline is provided in Figure \ref{fig:sfd_sfi}.

The SFI problem can be formalized as the following multi-hypothesis testing problem \cite{kay1993fundamentals} by exploding the \textit{at least one sensor fault} hypothesis $\mathcal{H}_1$ into multiple $\mathcal{S} = 2^S -1$ hypotheses:

\begin{equation} \label{eq:multi_hypo}
\begin{split}
\mathcal{H}_{11}:  \textit{ }  r^1_t \approx \textit{large}, r^2_t \approx \textit{small},\ldots, r^S_t \approx \textit{small} 
\\
\mathcal{H}_{12}:  \textit{ }  r^1_t \approx \textit{small}, r^2_t \approx \textit{large},\ldots, r^S_t \approx \textit{small} 
\\
\vdots \qquad \qquad \qquad \qquad \qquad \qquad \qquad \qquad \qquad
\\
\mathcal{H}_{1\mathcal{S}}:  \textit{ }  r^1_t \approx \textit{large}, r^2_t \approx \textit{large},\ldots, r^S_t \approx \textit{large} 
\\
\\
 \text{ for } t=t_I,\ldots,t_I+L \qquad \qquad \qquad \qquad \qquad 
\end{split}
\end{equation}
Clearly the hypothesis $\mathcal{H}_{11}$ refers to the single fault in sensor $1$ and so on, with $\mathcal{H}_{1\mathcal{S}}$ denoting the hypothesis of having all $S$ faulty sensors.
Within this multi-hypotheses framework, the optimal strategy for the SFI unit, (optimal in the maximum likelihood sense \cite{kay1993fundamentals}) would be to identify the faulty sensors resulting the in largest likelihood. \footnote{Alternatively, if a prior domain knowledge about some sensors requiring more attention than others, an optimal Bayesian approach could be used\cite{kay1993fundamentals}.}
The main issue of the optimal approach\footnote{In principle one could learn the PDF's $\{f(r_{t}^{s}|H_{1j})\}_{j=1}^{\mathcal{S}}$ from a subset of the data set $\{\bm{x}_t\}_{t=t_{I}}^{t_{I} + L}$, and then compute the likelihood function for all the hypotheses $\{H_{1j}\}_{j=1}^{\mathcal{S}}$ .} is related to its computational complexity of $\mathcal{O}(2^S)$ making the SFI a NP-hard problem \cite{van1990handbook}.

In order to make the problem more tractable, one could compute the \textit{contribution} of each sensor to the total residual \cite{alcala2009reconstruction}, \cite{dunia1996identification}. 
The Contribution Score $CS$ for a sensor $s$ over $M$ samples can be calculated as follows: \footnote{We omit the dependency on M for simplicity of notation.}
\begin{equation}
CS_{SFI}(s)=\frac{\sum_{t=t_I}^{t_I + L}r^s_n}{\sum_{s=1}^{S}\sum_{t=t_I}^{t_I + L}r^s_n}
\label{eq:sfi_contrib}
\end{equation}
For example, in the single sensor fault, the sensor with the largest contribution score is likely to be the faulty one, as formally written in the following:
\begin{equation}
\hat{s} = \argmax_{s=1,\ldots, S} CS_{SFI}(s)
\label{eq:sfi_contrib}
\end{equation}
For the multiple faults case, the subset of faulty sensors is given by the $S^*$ largest entries in the vector $\bm{CS_{SFI}}=[CS_{SFI}(1), \ldots, CS_{SFI}(S)]$.
One undesirable feature of this naive method is the fact that the cardinality $S^*$ of the faulty sensors subset is typically unknown.
In Section \ref{sec:greedyApproach}, we will introduce a greedy approach with worst-case computational complexity of $\mathcal{O}(S)$ for the identification of the multiple sensor faults, approach that does not require the prior knowledge of the number of faulty sensors.

\subsection{Learning with disentanglement}
\label{sec:disent}
During the training phase of the RNN, certain correlations between sensor outputs are picked up and are exploited to increase the accuracy of the prediction.
This is a desirable effect as long as there are no faults present.
As soon as there is a fault in one sensor, the error can flow into the predictions of other sensors obfuscating the true faulty sensor by reducing its contribution score $CS_{SFI}$. This \textit{smearing-out} effect was first studied in \cite{westerhuis2000generalized} and \cite{alcala2009reconstruction} and recently analyzed in \cite{van2013contribution}. 
A clear representation of the \textit{smearing-out} effect is given in the left plot of Figure~\ref{fig:contribPlot} where the contribution scores $CS_{SFI}$ for $S=8$ sensors are plotted against the increasing fault signature intentionally injected in the purple sensor.
One would expect that the $CS_{SFI}$ for the faulty sensor being the largest one. However, this holds only when the fault signature starts to be significantly large. The contribution plot will miss-classify the faulty purple sensor with the blue sensor for fault signature smaller than $15\%$ threshold.

To circumvent this problem we take inspiration from a recent work in \cite{kageback2018disentangled} where a novel regularization term, based on the covariance of the activation in a neural network layer, is added to the classical loss function. Through this regularization, the objective is to penalize the cross-correlation between the dimensions of the learned representations and, by doing it, helping artificial neural networks learn disentangled representations \cite{bengio2013representation}.
Similarly, in this paper we propose a disentangled RNN model with the 
introduction of a regularizer based on the covariance matrix between the RNN predictions $\bm{\hat{x}}_{t}$.
With this approach the goal is to learn a disentangled RNN that decreases the likelihood of having a sensor fault leaking into another sensor by reducing the weights of the network connections that are the ultimate responsible for the smearing-out effect.
This is achieved by penalizing the correlation between the sensors predictions (which are the last layer of the model) during the learning phase. 
Mathematically we define the new regularization term as:
\begin{equation}
\mathcal{L}_{\bm{C}}= \frac{1}{S^2} \|\bm{C}\|_1
\label{eq:disent}
\end{equation}
with $\|\bm{C}\|_1$ defined as:
\begin{equation}
\|\bm{C}\|_1 = \sum\limits^S_{i,j=1} | \bm{C}_{i,j}|
\label{eq:covNorm}
\end{equation}
is the element-wise L1 matrix norm of $\bm{C}$, where $\bm{C}$ is the covariance of the predictions, computed over the $N$ training samples:
\begin{equation}
\bm{C} = \frac{1}{N-1} \sum\limits^N_{i=1} (\bm{\Hat{X}} - \bm{\Hat{x}}_{avg} \bm{1}_N ) (\bm{\Hat{X}} - \bm{\Hat{x}}_{avg} \bm{1}_N ) ^\top
\label{eq:covMat}
\end{equation}
where $\bm{\Hat{X}}=[\bm{\Hat{x}}_1,\ldots,\bm{\Hat{x}}_N] \in \mathbb{R}^{S\times N}$ is the matrix of all the predictions, $\bm{1}_N$ is a $N$-dimensional row vector of all ones and $\bm{\Hat{x}}_{avg} \in \mathbb{R}^{S\times 1}$ is the average of $\bm{\Hat{X}}$ over the sample dimension \footnote{The vector $\bm{\Hat{x}}_{avg}$ is defined as  $\bm{\Hat{x}}_{avg}=[\Hat{x}^{1}_{avg},\ldots,\Hat{x}^{S}_{avg}]$, where the generic element $\Hat{x}^{s}_{avg} = \frac{1}{N} \sum\limits^N_{i=1} \Hat{x}^{s}_{i}$.}
The new loss function used during the training process becomes:
\begin{equation}
\mathcal{L}_{tot} = \mathcal{L}_{MSE} + \lambda\mathcal{L}_{\bm{C}}
\label{eq:lossFinal}
\end{equation}
where $\lambda$ is the regularization parameter that controls the importance of the regularization term.

In the middle plot of Figure~\ref{fig:contribPlot} we can appreciate the benefit in using disentanglement regularization as the contribution score for the purple sensor (the faulty one) is larger than the others even for lower offset level. Learning with disentanglement helps the SFI being more accurate even when the fault signature is significantly small (the fault signature threshold decreased from $15\%$, when no disentanglement was enforced, to approximately $5\%$).

\subsection{Targeted disentanglement}
\label{sec:tarDisent}
If one is really interested in increasing the sensitivity of the contribution score for one particular sensor (e.g. the purple one in our toy example), a more aggressive regularization strategy could be devised.
Instead of imposing a regularization term as in (\ref{eq:disent}) that depends on all the entries of the covariance matrix, a \textit{targeted} regularization term, affecting only a specific sensor, can be defined as:. 

\begin{equation}
\mathcal{L}_{\bm{C}_s}= \frac{1}{S} \|\bm{C}_s\|_{1}
\label{eq:disentTarget}
\end{equation}
where $s$ is the index of the sensor to be targeted and
\begin{equation}
\|\bm{C}_s\|_1 = \sum\limits^S_{j=1} | \bm{C}_{s,j}|
\label{eq:covTarget}
\end{equation}
is the element-wise L1 vector norm of $\bm{C}_s$, which is the $s$-th row of the matrix $\bm{C}$ and $\bm{C}_{s,j}$ is entry in the $s$-th row and $j$-th column of $\bm{C}$.
This would enforce a weaker correlation between the targeted sensor and all the others with the consequent effect of increasing its sensitivity of being isolated at the cost of reducing the sensitivity for the remaining sensors.

The result of the aforementioned method is shown in the right plot of Figure~\ref{fig:contribPlot}.
The faulty sensor (the purple one) is correctly identified with a very large sensitivity, as it can be acknowledged by observing the purple contribution score being the largest one over all the offset level values.

Such a technique, if on the one hand leads to a larger sensitivity for the targeted sensor, on the other might make the overall SFI performances worse (other sensors might be miss-classified due to an aggressive disentanglement applied to one sensors and not the others).
Therefore, the targeted disentanglement (it can be easily extended to a sub-set of sensors) must be used with caution in  contexts where prior information about the most \textit{critical} sensors\footnote{the adjective \textit{critical} might be applied to either sensors that are more prone to failure or sensors whose health status is critical for the system operations.} will drive the design of the regularization term.

\section{A greedy approach to the SFI problem}
\label{sec:greedyApproach}

SFI problem has combinatorial complexity \cite{baygun1995optimal} when multiple faults can occur concurrently as it was also discussed in Section \ref{sec:SFI}.
A $\ell_1$ regularized least-squares optimization with a $\ell_1$ penalty for the sensors fault (i.e. a bias) vector $\bm{\Delta}$ was introduced in \cite{gorinevsky2009estimation}. 
Within our framework, the approach based on sparsity can be mathematically formulated as follows:

\begin{equation}
\begin{split}
    \bm{\Hat{\Delta}_{\eta}} = \argmin_{\bm{\Delta}} \|   \bm{\hat{x}}- \bm{x}  - \bm{\Delta}\|^2_2  + \eta \|\bm{\Delta}\|_1
\end{split}
\label{lab:e_optimisation}
\end{equation}
where, to ease the notation, we denote with $\bm{x} = [\bm{x}_{t_I}, \dots, \bm{x}_{t_I+L}]$ and $\bm{\hat{x}} = [\bm{\hat{x}}_{t_I}, \dots, \bm{\hat{x}}_{t_I+L}]$ the sensors measurements and their predictions over the SFI integration time, respectively.

It is however not clear how to properly select the regularization parameter $\eta$ to mitigate the risk of incurring in miss-classification of the faulty sensors \cite{li2014sensor}.
The authors in \cite{li2014sensor}, introduced an algorithm to determine the regularization parameter automatically from the data based on a bootstrap approach, also known as BINCO method  \cite{li2013bootstrap}.
The BINCO method is quite complex to implement. In addition, by also requiring some tuning parameters, it goes against the same principle it was introduced for in the first place. 

In \cite{zymnis2009relaxed} the authors formulated the SFI problem as a maximum a posteriori probability (MAP) estimation problem \cite{kay1993fundamentals}, and solved it using a convex relaxation followed by rounding, and, possibly, local optimization. 
For this elegant approach to work, however, both a measurement model for the sensor and a prior probability of the sensors fault vector are needed.

Because of the aforementioned problems related to the approaches using sparsity, in this work we introduce a greedy algorithm (that we call GreedyIso) that uses only the assumption on the additive bias model as specified in Equation (\ref{eq:sensor_fault}). Another important feature of the GreedyIso is that its performance is not affected by the regularization parameter as in \cite{gorinevsky2009estimation}. 
Its worst-case computational complexity scales linearly with the number of sensors, that is $\mathcal{O}(S)$.
Before elaborating on the GreedyIso algorithm, which is largely-detailed in the pseudo-code given in the Algorithm \ref{alg:MFgreedy}, a few definitions are needed.

We denote with $\Bar{R} = \frac{1}{L}\sum\limits_{t=t_I}^{t_I+L} R_t$ the average residual norm over the SFI integration time.
Given a list $\bm{a} \subseteq [1,\dots,S]$ of length $Q \leq S$ and a vector $\bm{x}$ of length $S$, we define the access operator as follows:

\begin{equation} \label{eq:access}
\bm{x}^{\bm{a}} = [x^{a_1},\ldots,x^{a_Q}]^\top.\\
\end{equation}


\begin{algorithm}
\caption{GreedyIso}\label{alg:MFgreedy}
\begin{algorithmic}[1]
\State $\text{Using } \bm{x} \text{ compute } \Bar{R} \text{ and } P_{D}, \text{ from eq. (\ref{lab:Pdetect})}$
\State $\bm{\mathrm{f}} \gets [\ ], s = 1$
\While{$P_{D} > 0 \text{ OR } s\leq S$}{}
\State $\bm{y} \gets \bm{x}  $
\State $s_{max} = \argmax(\bm{CS_{SFI}}) $
\State $\bm{CS_{SFI}}[s_{max}] = -1$
\State $\bm{\mathrm{f}}.append(s_{max})$
\For{$t = t^* \dots t_I+L $}
\State  $\bm{\hat{y}}_{t} = \bm{g^{\bm{\theta}}}(\bm{\hat{y}}_{t-1}, \bm{y}_{t-1})$
\State  $\bm{y}_{t}^{\bm{\mathrm{f}}} = \bm{\hat{y}}_{t}^{\bm{\mathrm{f}}}$
\EndFor
\State $\bm{\Hat{\Delta}^\bm{\mathrm{f}}} = \frac{1}{L}\sum\limits_{t=t_I}^{t_I+L}\bm{x}_t^{\bm{\mathrm{f}}}-\bm{y}_{t}^{\bm{\mathrm{f}}}$
\State $\bm{x_{new}^\bm{\mathrm{f}}} \gets \bm{x}^{\bm{\mathrm{f}}} - \bm{\Hat{\Delta}^\bm{\mathrm{f}}}$
\State $ \text{Using }\bm{x}=[\bm{x}_{new}^{\bm{\mathrm{f}}},\bm{x}^{\bm{\mathrm{f}}^c}]\text{ compute }\Bar{R}\text{ and }P_{D}$
\State $ \text{ to get }P_{D,new} \text{ and } \Bar{R}_{new}$
\If {$P_{D,new} \leq P_{D}\ \bm{and} \ \Bar{R}_{new} < \Bar{R}$}
\State    $P_{D} = P_{D,new}$
\State    $\Bar{R} = \Bar{R}_{new}$
\Else 
\State    $\bm{delete}\ \bm{\mathrm{f}}[-1]$
\EndIf
\State $s = s + 1$
\EndWhile
\end{algorithmic}
\end{algorithm}
The GreedyIso algorithm starts from the sensor with the largest contribution score as per line 5 and add it to the list $\bm{\mathrm{f}}$ of faulty sensors. The predicted signal evaluated only for the faulty sensors in $\bm{\mathrm{f}}$, is computed as per line 9. Then an estimate  $\bm{\Hat{\Delta}}$ of the sensors bias vector $\bm{\Delta}$ is computed by averaging the difference between the predicted and observed signal over the SFI integration time interval as in line 11. 
The reason why the loop in line 8 is extended over the interval that goes from $t^*$ up to $t_{I} + L$, is to improve the estimate of the bias vector $\bm{\Delta}$ by feeding the RNN model with non-faulty sensors measurements prior to $t^*$. 

The $\bm{\Hat{\Delta}}$ is subtracted from the observed signal to correct for the faulty sensors as in line 12. The parameters $P_D$ and $\Bar{R}$ (using the corrected signal) combined with the signal for the non-faulty sensors, are re-computed as in line 13\footnote{$\bm{\mathrm{f}}^c$ is defined as the complement of $\bm{\mathrm{f}}$, that is the list of non-faulty candidates.}. By correcting the faulty sensors values, one would expect that $P_D$ and $\Bar{R}$ will decrease as the effect of the faults diminishes.\\
If $P_D$ decreases or stays constant(the latter could occur in the case of multiple faults) and the residual decreases, the current fault candidate remains in the fault list $\bm{\mathrm{f}}$ and the thresholds are updated as per lines 16 and 17. Otherwise, the candidate is removed from $\bm{\mathrm{f}}$ as per line 19.
This is repeated until either $P_{D} = 0$ or all sensors have been processed at least once (the condition in line 6, ensures that for each round of the algorithm a new potential faulty sensor added to the list $\bm{\mathrm{f}}$).
From the logical condition in the \textbf{while} loop as per line 3, we can determine that the worst-case complexity (the algorithm might stop even before going through all the $S$ sensors) is of the order of $\mathcal{O}(S)$.
The output of the GreedyIso algorithm is the list $\bm{\mathrm{f}}$ of faulty sensors. As a by-product of the algorithm, the estimates of the different biases at the faulty sensors are given in the vector $\bm{\Hat{\Delta}^\bm{\mathrm{f}}}$. 

One of the key components of the GreedyIso algorithm is the contribution score vector $\bm{CS_{SFI}}$ that regulates the greedy process of constructing the fault list $\bm{\mathrm{f}}$ deciding which potential faulty sensor with the largest score, is first processed (see line 7 in Algorithm \ref{alg:MFgreedy}) and promoting the sensor with the second largest score in the next iteration and so forth.
A variant of the GreedyIso algorithm is obtained by coupling the approach as in (\ref{lab:e_optimisation}) with the GreedyIso algorithm. It is  named GreedyIsoSparse and its description is given in the Algorithm \ref{alg:MFgreedy_sparse} where the only difference with the Algorithm \ref{alg:MFgreedy} is that the contribution score vector $\bm{CS_{SFI}}$ is replaced with the vector of the absolute value of all the elements in $\bm{\Hat{\Delta}_{\eta}}$, denoted with $\mathrm{abs}(\bm{\Hat{\Delta}_{\eta}})$, where $\mathrm{abs}$ acts element-wise on each element of $\bm{\Hat{\Delta}_{\eta}}$.
The gist of the GreedyIsoSparse algorithm is that the faulty sensors will result in the largest (in absolute value) elements of $\bm{\Hat{\Delta}_{\eta}}$, while the healthy sensors will have biases close to zero.
The GreedyIsoSparse algorithm depends on the regularization parameter $\eta$ for which different tuning might lead to contrasting fault list results.
Lastly, GreedyIsoSparse algorithm might appear somehow convoluted because of the fact that the bias vector $\bm{\Hat{\Delta}}$ is estimated twice in the procedure. The first estimate is given by $\bm{\Hat{\Delta}_{\eta}}$, while the second is given in line 11. However, it is important to reinforce that $\bm{\Hat{\Delta}_{\eta}}$ is used only to drive the selection process of the potential faulty sensors into the following steps of the algorithm.

\begin{algorithm}
\caption{GreedyIsoSparse}\label{alg:MFgreedy_sparse}
\begin{algorithmic}[1]
\State $\text{Using } \bm{x} \text{ compute } \Bar{R} \text{ and } P_{D}, \text{ from eq. (\ref{lab:Pdetect})}$
\State $\bm{\mathrm{f}} \gets [\ ], s = 1, \bm{\Hat{\Delta}}_{\mathrm{abs}}=\mathrm{abs}(\bm{\Hat{\Delta}_{\eta}})
$
\While{$P_{D} > 0 \text{ OR } s\leq S$}{}
\State $\bm{y} \gets \bm{x}  $
\State $s_{max} = \argmax(\bm{\Hat{\Delta}}_{\mathrm{abs}}) $
\State $\bm{\Hat{\Delta}}_{\mathrm{abs}}[s_{max}] = -1$
\State $\bm{\mathrm{f}}.append(s_{max})$
\For{$t = t^* \dots t_I+L $}
\State  $\bm{\hat{y}}_{t} = \bm{g^{\bm{\theta}}}(\bm{\hat{y}}_{t-1}, \bm{y}_{t-1})$
\State  $\bm{y}_{t}^{\bm{\mathrm{f}}} = \bm{\hat{y}}_{t}^{\bm{\mathrm{f}}}$
\EndFor
\State $\bm{\Hat{\Delta}^\bm{\mathrm{f}}} = \frac{1}{L}\sum\limits_{t=t_I}^{t_I+L}\bm{x}_t^{\bm{\mathrm{f}}}-\bm{y}_{t}^{\bm{\mathrm{f}}}$
\State $\bm{x_{new}^\bm{\mathrm{f}}} \gets \bm{x}^{\bm{\mathrm{f}}} - \bm{\Hat{\Delta}^\bm{\mathrm{f}}}$
\State $ \text{Using } \bm{x}=[\bm{x}_{new}^{\bm{\mathrm{f}}},\bm{x}^{\bm{\mathrm{f}}^c}] \text{ compute } \Bar{R} \text{ and } P_{D}$
\State $ \text{ to get }P_{D,new} \text{ and } \Bar{R}_{new}$
\If {$P_{D,new} \leq P_{D}\ \bm{and} \ \Bar{R}_{new} < \Bar{R}$}
\State    $P_{D} = P_{D,new}$
\State    $\Bar{R} = \Bar{R}_{new}$
\Else 
\State    $\bm{delete}\ \bm{\mathrm{f}}[-1]$
\EndIf
\State $s = s + 1$
\EndWhile
\end{algorithmic}
\end{algorithm}



\begin{center}
\begin{figure}
\includegraphics[width=1\linewidth]{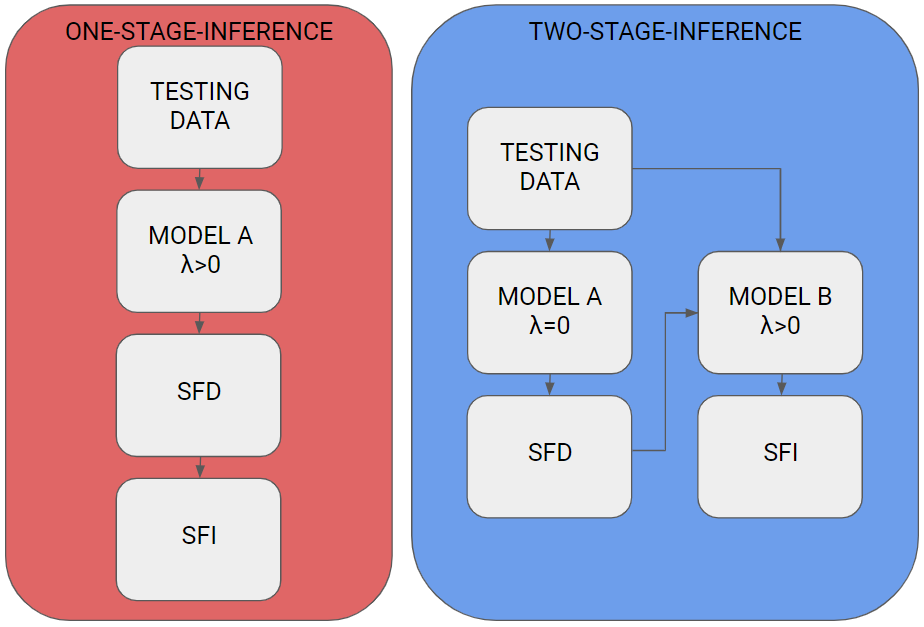}
   \caption{Comparison diagram for one-stage vs two-stage-inference. Shown in red is the inference process for one single RNN model with disentanglement used for SFD and SFI. In blue one can see a two models set-up, where one model trained without disentanglement is used for SFD and the other trained with disentanglement is used for SFI.}
\label{fig:blockDiagram}
\end{figure}
\end{center}

\section{SFD-SFI: Two-stage architecture}
\label{sec:two_stage}

In Section \ref{sec:disent} we introduced the disentanglement model with the objective of improving the SFI by reducing the smearing-out
effect.
The severity of the regularization can be controlled by the regularization parameter $\lambda$. Larger $\lambda$ enforces a stronger constraint on the RNN to have uncorrelated output. This, unfortunately, has a negative impact on the RNN's ability to predict, as the term $\bm{\mathcal{L}}_{MSE}$ has a lower contribution than that of $\lambda\mathcal{L}_{\bm{C}}$ to the total loss $\mathcal{L}_{tot}$. This will result in a more "noisy" residual $R_t$ with the subsequent SFD performance degradation.
One possible strategy, is to design a two-stage architecture that consists of training train two RNN models as depicted in Figure~\ref{fig:blockDiagram}. One model with $\lambda=0$ that is used exclusively for SFD, where more accurate predictions are required for the residuals. While for the second model we set $\lambda>0$ and we only use this model for the SFI process, where we want as little \textit{smearing-out} as possible. This allows for the best of both worlds at the cost of having to train two models.

\begin{center}
\begin{figure}
\includegraphics[width=1\linewidth]{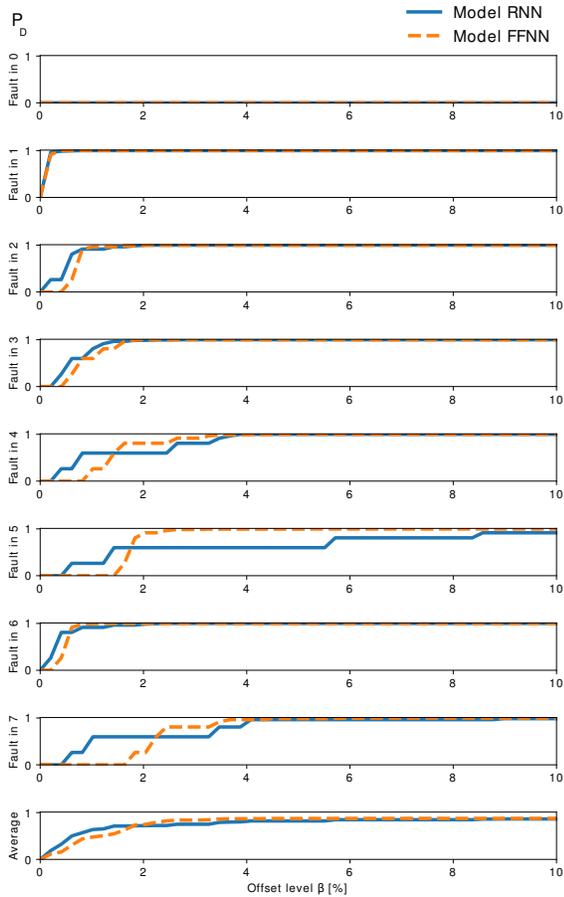}
   \caption{Probability of detection $P_D$ as function of the offset level $\beta$ for the two models, which are the RNN and the FFNN. The faster the detection rate $P_D$ goes to 1, the better the model is.}
\label{fig:fairComp}
\end{figure}
\end{center}

\section{Numerical Results}
\label{sec:num_results}

In this section we empirically verify the performances of the SFD-SFI algorithms that have been introduced in the previous sections. We consider a data set generated by a sensor system with 8 sensors monitoring a real petrochemical plant.
The data set consists of 185444 time ordered observations from 8 different sensors.
The data set is split into a training set, validation set and testing set of size 122641, 13627 and 49176 respectively. 
Sensor faults have been injected into the faultless training data set according to the model as described in Equation (\ref{eq:sensor_fault}).
All models used in this section are trained with TensorFlow\footnote{https://www.tensorflow.org/}.
The values for all the parameters used in this section are given in Table \ref{tab:1}.



\begin{table}[]
\centering
\begin{tabular}{|l|l|}
\hline
\multicolumn{2}{|c|}{\textbf{Model and Algorithm parameters}}                                                                                                                                                                                         \\ \hline
\multicolumn{1}{|c|}{\textbf{RNN}}              & \multicolumn{1}{c|}{\textbf{FFNN}}                                                                                                                                                                  \\ \hline
Number of epochs = 8                            & Number of epochs = 8                                                                                                                                                                                \\ \hline
Batchsize = 110                                 & Batchsize = 110                                                                                                                                                                                     \\ \hline
Hidden Units in GRU = 32                        & \begin{tabular}[c]{@{}l@{}}Number of Hidden Layers = 1\\ Number of Hidden Units = 30\end{tabular}                                                                                                   \\ \hline
                                                & Windowsize $w$ = 8                                                                                                                                                                                  \\ \hline
Solver = Adam                                   & Solver = Adam                                                                                                                                                                                       \\ \hline
Learning Rate = 0.001                           & Learning Rate = 0.001                                                                                                                                                                               \\ \hline
Activation Function = Tanh                      & Activation Function = Sigmoid                                                                                                                                                                       \\ \hline
Disentanglement Regularization $\lambda$ = 0.01 &                                                                                                                                                                                                     \\ \hline
\multicolumn{1}{|c|}{\textbf{SFD}}              & \multicolumn{1}{c|}{\textbf{SFI}}                                                                                                                                                                   \\ \hline
$P_{fa}$=0.1                                    & L=60                                                                                                                                                                                                \\ \hline
$p_{fa}$=0.01                                   & \begin{tabular}[c]{@{}l@{}}Parameters to solve equation (\ref{lab:e_optimisation})\\ Solver = Adam\\ Learning Rate = 0.1\\ Number of iterations = 30\end{tabular} \\ \cline{1-1}
M=60                                            &                                                                                                                                                                                                     \\ \hline
\end{tabular}
\caption{Table listing all the parameters used to generate the numerical results of this section.}
\label{tab:1}
\end{table}

\begin{center}
\begin{figure}
\includegraphics[width=1\linewidth]{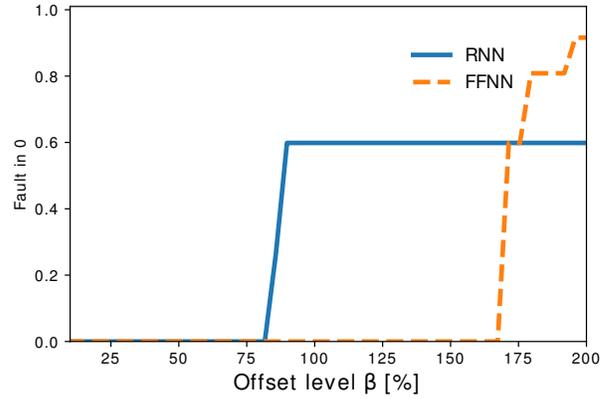}
   \caption{$P_D$ as a function of the offset level $\beta$ in the extended range for the fault injected into sensor 0.}
\label{fig:sensor0}
\end{figure}
\end{center}

\subsection{Performance measures}

The three performance statistical measures used in this section for the evaluation of the proposed algorithms are:
\begin{itemize}
    \item The probability of detection $P_D$ for a given fixed $P_{fa}$, which is pre-specified by the user.
     \item The accuracy for the single fault, defined  as 
     \begin{equation}
        ACC = \frac{Number\ of\ correct\ isolations}{Total\ number\ of\ isolations}
        \label{eq:acc}
     \end{equation}
      \item The mean intersection over union  (mIoU)~\cite{levandowsky1971distance, intoverunion, gouk2016learning} for the multiple faults case, where the mean is computed over the total number of runs of the numerical experiment.
\end{itemize}
Given two sets $A$ and $B$, the intersection over union (IoU), also known as Jaccardi index \cite{jaccard1901distribution}, is defined as:

\begin{equation}
IoU = \frac{|A \cap B |}{|A \cup B |}
\label{eq:iou}
\end{equation}
Unlike the single fault case classification, there is no single appropriate definition for accuracy when
performing SFI for multiple faults. This leads us to the choice of the IoU distance as opposed to other similarity functions, such as the Hamming distance \cite{zhang2013review} for instance, for which the sets A and B have to share the same cardinality.
But the output of the SFI, that is a set of the predicted faulty sensors, does not necessarily have the same size of the set of the true faulty sensors. 
In this respect, the IoU is a reasonable choice as it captures the proportion of
correctly predicted faulty sensors out of the potential faulty sensors (predicted faulty sensors and actually true faulty sensors).

\subsection{RNN vs. FFNN}
In this first experiment, we want to demonstrate the performance of the RNN over the FFNN. For this, we compare the detection sensitivity for both models over increasing offset level $\beta$. To have a fair comparison, the number of hidden layers are adjusted, such that both networks have the same amount of capacity.
In Figure \ref{fig:fairComp} we present 9 plots showing $P_D$ as function of $\beta$ for when the fault is injected one at a time, in sensor number 0, sensor number 1, up to sensor number 7. The last plot, which is the average of the first 8 plots, is a proxy of the average SFD behavior over the different faulty cases.
For instances, for the case where the fault is injected in sensor 7, the SFD, based upon the RNN model, reacts to the increasing offset faster than that based on the FFNN.
However, as the average $P_D$ in the last plot suggests, it appears that the performances of the SFD for the two models are very close to each other. 
This comparison is performed to make our analysis more complete by showing that other models than the RNN, can be conceived. As mentioned in the Introduction section, the main ideas of this paper apply regardless of the prediction model used to generate the residuals.

Last but not least, it is worth to note that for this particular data set, the fault injected in sensor 0 is undetectable in the chosen range of the offset level $\beta$, as confirmed by the first plot showing a zero probability of detection for all the values of $\beta$. 
This might be due to the fact that the fault signature is buried in the inherent complex dynamics of the measurements in sensor 0. 
By extending the range of $\beta$, the detector starts to kick in, as you can see in Figure \ref{fig:sensor0}.

\subsection{One vs. Two Stage Architecture}
From this section onward the results were generated using the RNN model. Here we assess the SFD performance in terms of probability of detection for the one-stage vs the two-stage architecture discussed in Section \ref{sec:two_stage}.

\begin{center}
\begin{figure}
\includegraphics[width=1\linewidth]{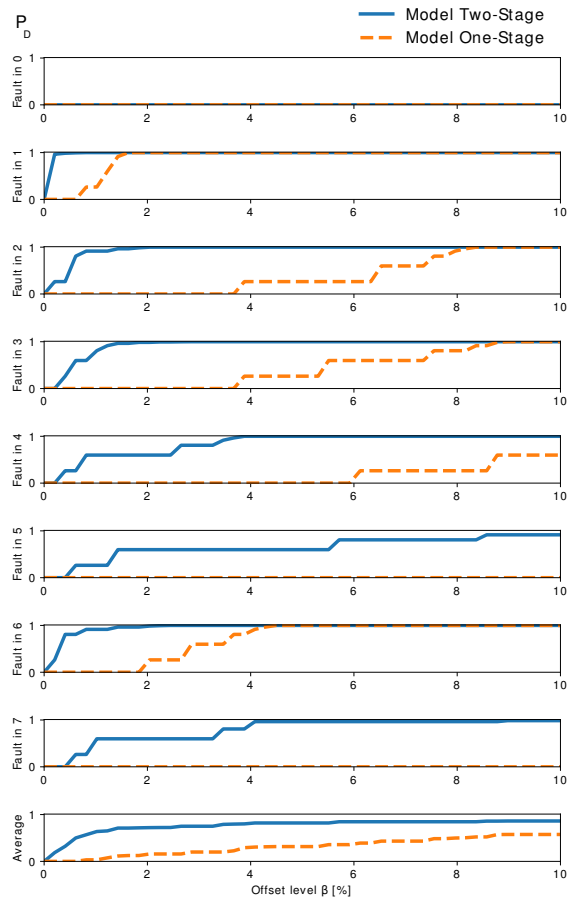}
   \caption{Probability of detection $P_D$ as function of the offset level $\beta$ for the the one-stage and two-stage architectures. The faster the detection rate $P_D$ goes to 1, the better the model is.}
\label{fig:onevstwo}
\end{figure}
\end{center}

The resulting plots can be seen in Figure \ref{fig:onevstwo}. In all cases the detector for the two-stage architecture reacts much faster than the one-stage architecture. The main cause for this behavior is due to the disentanglement regularization term, because of which the network is discouraged from mixing information from the different sensors. While this leads to better results for SFI tasks, it has a negative effect on the detection capabilities of the SFD system.
By having two different prediction models (one for the SFD and one regularized model for the SFI), the two-stage architecture does not suffer from the above pathological problem.

\subsection{Disentanglement vs. No Disentanglement for the single sensor fault}
\label{sec:testDis}
In this section we examine the effectiveness of the disentanglement for the single sensor fault case.
The contributions scores are computed for the two-stage architecture and a modified one-stage architecture where the model has been trained without the application of the disentanglement (i.e., $\lambda=0$), due to the fact that the detection capabilities would be greatly impaired with disentanglement as shown in the previous experiment. 
The accuracy $ACC$ for both the two architectures is computed over 100 runs where at each run a random offset, which is uniformly distributed in the range [5\%, 30\%], is added into one randomly chosen sensor.
Since we are dealing with a single sensor fault, we simply predict the faulty sensor to be the one with the largest contribution score. In principle, we could have used either the GreedyIso algorithm or the GreedyIsoSparse for the same task, but our objective in this experiment is to quantify the benefit of the disentanglement regardless of the algorithm used for isolating the faulty sensor. Decoupling the disentanglement effect from the GreedyIso (or its variant), allows us to characterize the disentanglement approach at its full power. 
These results are shown in the first column of Table \ref{tab:multifault}, where the accuracy for the two-stage with disentanglement is significantly larger than the one for the case where no disentanglement was used.
This numerical results is very interesting and it confirms the intuition that was highlighted in the toy example in Section \ref{sec:disent}.
However, it should be recalled that fault signatures in the range [5\%, 30\%] injected into sensor 0 are not detectable. This leads to an automatic loss of accuracy of $\frac{1}{8}$ such that the best attainable accuracy is $87.5\%$, which results in a corresponding accuracy of approximately $95\%$.
 
\subsection{Performance for multiple faults}
\label{sec:testMulti}

In this final experiment, we investigate the performance of the GreedyIso and GreedyIsoSparse algorithms for the multiple sensor fault case within the two-stage architecture.
Because of multiple faults, we need to use another performance metric than accuracy. Instead we use the mIoU, defined as follows:

\begin{equation}
mIoU = \frac{1}{N_{run}}\sum_{i=1}^{N_{run}} IoU(\bm{\mathrm{f}}_i,\bm{\mathrm{f}}^{true}_i)
\label{eq:miou}
\end{equation}

where $\bm{\mathrm{f}}_i$, $\bm{\mathrm{f}}^{true}_i$ and $N_{run}$ are the set of faulty sensors predicted by the GreedyIso algorithm, the set of true faulty sensors, and the number of runs, respectively.
We compute the $mIoU$ over 100 runs, where we randomly pick an offset level in the range [5\%, 30\%] and insert the offset into two or three evenly randomly picked sensors from all sensors.

In the second column of Table \ref{tab:multifault} we show the resulting performances in terms of mIoU.

The Table \ref{tab:multifault} shows that the GreedyIso  is $83.3\%$.
However, as previously mentioned, fault signatures in the range [5\%, 30\%] injected into sensor 0 are not detectable. This lead to an automatic loss of mIoU of $12.5\%$ such that the best attainable mIoU is $87.5\%$, which results in an actual mIoU of approximately $95.2\%$.
The GreedyIso shows superior performance than the GreedyIsoSparse for any value of $\eta$.
The explanation for this might reside in the fact that the contribution score vector $\bm{CS_{SFI}}$, which is  used by the GreedyIso algorithm to guide the process of building up the fault list $\bm{\mathrm{f}}$, contains more useful information than its counterpart $\bm{\Hat{\Delta}}_{\mathrm{abs}}$ used in the GreedyIsoSparse approach.

The mIoU of the GreedyIsoSparse algorithm decreases dramatically for large values of $\eta$ due to a very strong regularization suppressing the main loss term $\|   \bm{\hat{x}}- \bm{x}  - \bm{\Delta}\|^2_2$ as in the  equation (\ref{lab:e_optimisation}).
That said, for a significant large range of values for $\eta$, the GreedyIsoSparse achieves consistent results suggesting that it is not sensitive to the different choice of the regularization parameter $\eta$. 

Lastly, we would like to remark that we intentionally decided not to include the results for the 
plain sparse solution of the optimization in (\ref{lab:e_optimisation}), since we still need to apply a threshold to the resulting $\bm{\Hat{\Delta}_{\eta}}$ vector by forcing the smallest values of the vector to zero, for instance.
But this thresholding procedure is arbitrary.

\begin{table}[h]
\centering
\begin{tabular}{|c|c|c|}
\hline
\textbf{Methods}                                                 & \textbf{\begin{tabular}[c]{@{}c@{}}Single Fault\\ \textit{ACC}\end{tabular}} &  \textbf{\begin{tabular}[c]{@{}c@{}}Multiple Faults\\ mIoU\end{tabular}}\\ \hline
\begin{tabular}[c]{@{}c@{}c@{}}\textbf{GreedyIso}\\ \end{tabular}  & 83\%                                                                                                                                   & 83.3\% \\ \hline
\begin{tabular}[c]{@{}c@{}c@{}c@{}}\textbf{GreedyIsoSparse}\\    $\eta=0$ \end{tabular}  & 83\%                                                                                                                                   & 80.3\% \\ \hline
\begin{tabular}[c]{@{}c@{}c@{}c@{}}\textbf{GreedyIsoSparse}\\    $\eta=0.01$ \end{tabular}  & 83\%                                                                                                                                   & 80.3\% \\ \hline
\begin{tabular}[c]{@{}c@{}c@{}c@{}}\textbf{GreedyIsoSparse}\\    $\eta=1$ \end{tabular}  & 83\%                                                                                                                                   & 80.3\% \\ \hline
\begin{tabular}[c]{@{}c@{}c@{}c@{}}\textbf{GreedyIsoSparse}\\    $\eta=10$ \end{tabular}  & 83\%                                                                                                                                   & 80.3\% \\ \hline
\begin{tabular}[c]{@{}c@{}c@{}c@{}}\textbf{GreedyIsoSparse}\\    $\eta=100$ \end{tabular}  & 76\%                                                                                                                                   & 80.3\% \\ \hline
\begin{tabular}[c]{@{}c@{}c@{}c@{}}\textbf{GreedyIsoSparse}\\    $\eta=500$ \end{tabular}  & 56\%                                                                                                                                   & 73.1\% \\ \hline
\begin{tabular}[c]{@{}c@{}c@{}c@{}}\textbf{GreedyIsoSparse}\\    $\eta=1000$ \end{tabular}  & 43\%                                                                                                                                   & 65.8\% \\ \hline
\end{tabular}
\caption{}
\label{tab:multifault}
\end{table}

\section{Conclusions}
\label{sec:conclusions}
In this paper we introduced a novel approach to mitigate the \textit{smearing-out} effect that affects the performance of any SFI algorithm based on residual analysis. Such an approach relies on adding disentanglement regularization term to the original loss function of the RNN. In the experiments we showed that the disentangled RNNs have promising performance for SFI and generate insights for future investigations.

Moreover, we proposed a scalable SFI algorithm with linear complexity in the number of sensors. A variant based on solving a sparse optimization problem was also introduced and evaluated.

Future works include exploring different sensor fault models (e.g. time-varying bias) and different disentanglement strategies capturing additional prior knowledge. It could also be of interest to characterize and analyze the performance of the SFD for challenging scenarios where the values of the bias are very low, using for instance the locally optimum detection paradigm \cite{kassam2012signal}. 

\ifCLASSOPTIONcaptionsoff
  \newpage
\fi



%
\printbibliography


%








\end{document}